\documentstyle[aps,prb,twocolumn]{revtex}
        \def\et{{\it et al.}}
\tightenlines

\begin{document}
\draft
\twocolumn[\hsize\textwidth\columnwidth\hsize\csname @twocolumnfalse\endcsname
%%%%%%%% draft format (above) %%%%%%%%%%

\title{Hydrogen-bonded Silica Gels Dispersed in a Smectic 
Liquid Crystal: A Random Field XY System}

\author{S. Park$^{1}$, R. L. Leheny$^{1,2}$, 
R. J. Birgeneau$^{1,3}$, J.-L. Gallani$^{1,}$\thanks{Permanent Address: 
IPCMS-GMO, Strasbourg, France.}, C. W. Garland$^{1}$, G. S. 
Iannacchione$^{4}$}

\address{$^{1}$Center for Materials Science and Engineering, 
Massachusetts Institute of Technology,
Cambridge, MA 02139, USA
\\
$^{2}$Department of Physics and Astronomy, Johns Hopkins
University,
Baltimore, MD 21218, USA
\\
$^{3}$Department of Physics, University of Toronto, Toronto, Canada 
M5S 1A1
\\
$^{4}$Department of Physics, Worcester Polytechnic Institute,
Worcester, MA 01609, USA
}

\date{\today}
\maketitle

\begin{abstract}
The effect on the nematic to smectic-A transition in octylcyanobiphenyl 
(8CB) due to 
dispersions of hydrogen-bonded 
silica (aerosil) particles is characterized with high-resolution x-ray scattering.  
The particles form weak gels in 8CB creating a quenched disorder 
that replaces the transition with the growth of short 
range smectic correlations.  The correlations include thermal critical fluctuations 
that dominate at high temperatures
and a second contribution that quantitatively matches the static fluctuations of a 
random field system and becomes important at low temperatures.

\end{abstract}
\

\vskip2pc
]
\narrowtext
%%%%% prl format (below) %%%%%%%%%%%%%%
%\phantom{.}
%]
%\narrowtext
%%%%%%%%% prl (above) %%%%%%%%%%%%%%%%%%

Quenched disorder in condensed matter systems can profoundly affect 
their thermodynamic behavior, creating unique challenges for statistical 
mechanics.  For fluids, confinement within a random porous medium offers a convenient 
method for introducing such disorder, and studies of superfluid helium 
~\cite{kim_He} and binary 
liquids~\cite{frisken_bl} in aerogels -- rigid, highly-porous chemically-bonded silica 
gels -- have 
led to insights into the implications of disorder on phase transitions.  
Similarly, the fragile nature of the mesomorphic phases of liquid crystals and their 
sensitivity to surface interactions make these systems well suited for studies within 
porous media.  In particular, the nematic (N) to smectic-A (SmA) 
transition, which typically exhibits anisotropic 3D-XY critical 
behavior, offers an ideal case for 
understanding the effects of a random environment on continuous 
symmetry fluctuations.
The demonstration through x-ray scattering~\cite{bellini-science,rappaport} and 
calorimetry~\cite{lei} of the destruction of the N-SmA transition by confinement 
in aerogels has shed light on these effects. 
Detailed theoretical work has further elucidated the fragility of the smectic phase 
to arbitrarily weak quenched disorder and has introduced the possibility that for sufficiently 
weak disorder novel ``Bragg glass'' phases may become stable~\cite{leo}.

In an effort to probe liquid crystals with increasingly weak disorder 
in a controlled way, we have conducted a high-resolution x-ray scattering 
study of the effect of hydrogen-bonded silica (aerosil) 
gels on the N-SmA
transition in octylcyanobiphenyl (8CB).  We find that the constraints imposed by these 
weakly-connected gels profoundly alter the 
smectic ordering, leading to behavior quantitatively like that induced 
by random fields.  Random field systems have been 
fruitful models for exploring theoretically  effects of quenched 
disorder, and random field Ising magnets have 
provided the primary experimental testing ground for these 
ideas~\cite{birgeneau}.  
8CB in porous silica thus introduces an opportunity to make experimental contact with random field 
theories for a transition that breaks 
a continuous symmetry~\cite{leo,imry-ma,aizenman,fisher1}.  Consistent 
with theoretical predictions, we find that the smectic phase is 
destroyed by the disorder and is replaced by the growth of short-range smectic 
correlations.  The low-temperature correlation length exceeds the typical 
void size in the gels by approximately a factor of four, consistent with 
the picture of the aerosil 
creating a weak to intermediate disorder.  The temperature 
dependence of the correlation function for different densities of gel further 
indicates that the disorder suppresses nematic fluctuations in the 
liquid crystal.

The smectic state is distinguished from the nematic by a one-dimensional
modulation in density, and a two-component order parameter 
specifying the amplitude 
and phase of the modulation characterizes the N-SmA transition.  To 
introduce quenched disorder, we disperse in the 8CB hydrophilic aerosil, 70 {\AA} 
diameter silica particles coated with hydroxyl groups.  At 
volume fractions above roughly 1\%, these particles form a 
hydrogen-bonded thixotropic gel, and interactions between the surfaces of the 
gel 
and liquid crystal molecules (``anchoring'') introduce locally preferred orientations and phases to the 
smectic ordering that compete with the globally uniform order of 
the pure smectic.  Aerosil gels have strong similarities with aerogels 
but represent a weaker disorder for the smectic in two essential respects.  
First, the ability to grow the aerosil gel directly 
within the liquid crystal permits structures of higher porosity.  
Also, as discussed below, for a given silica density the 
hydrogen-bonded aerosils perturb the smectic ordering less severely 
than rigid aerogels.  However, consistent with recent 
studies~\cite{germano,finotello} we find that aerosils above 1\% volume fraction 
create a reproducible, density-dependent 
quenched disorder.
In our experiments we have tuned the strength of 
this disorder by varying the density of silica, $\rho_{s}$, from 0.025 
grams of SiO$_{2}$ per cm$^{3}$ 
of 8CB, just 
above the gelation threshold, to 0.341 grams of SiO$_{2}$ per cm$^{3}$ of 8CB.  
Reference~\cite{germano} provides details of the sample preparation and 
fractal structure of aerosil gels.  

To characterize the effect of the aerosil gels on the N-SmA transition, we have measured with 
x-ray scattering the static structure factor, $S(q)$,   
which provides direct information about the correlation 
function of the smectic order.  The experiments were conducted on the X20 beamline of 
the National Synchrotron Light Source using 8 keV photons.  1 mm 
thick samples 
were held between kapton sheets in an oven containing dry nitrogen 
gas, and the scattering was performed in transmission geometry.  
Following prolonged exposure to the x-rays, the 8CB with dispersed aerosil 
displayed evidence of damage, 
including progressively degraded smectic order.  Therefore, we carefully limited the exposures to levels below which any 
damage was detectable.

Figure 1 shows representative x-ray scattering lineshapes for two 
values of $\rho_{s}$.  In accord with theoretical predictions of 
Radzihovsky, Toner, \et~\cite{leo,ward}, for all densities and for the full experimental temperature 
range, the diffraction peak corresponding to smectic layering is 
markedly broader than the resolution-limited peak measured for pure 8CB in the 
SmA phase, demonstrating the destruction 
of the smectic long range order by the aerosil.  We model this short-ranged order with a structure factor 
containing two contributions:
\begin{eqnarray}
    S({q}) = 
  \frac{\sigma_{1}}{1+(q_{\|}-q_{0})^{2}\xi_{\|}^{2}+q_{\bot}^{2}\xi_{\bot}^{2}+cq_{\bot}^{4}\xi_{\bot}^{4}} \nonumber \\
   +\frac{\sigma_{2}(\xi_{\|}\xi_{\bot}^{2})}{(1+(q_{\|}-q_{0})^{2}\xi_{\|}^{2}+q_{\bot}^{2}\xi_{\bot}^{2}+cq_{\bot}^{4}\xi_{\bot}^{4})^{2}}
\end{eqnarray} 
where the wave vector $q_{\|}$ is along the smectic layer normal and $q_{\bot}$ 
is perpendicular to it.

The first term in $S(q)$, an anisotropic Lorentzian with quartic corrections, 
is the form for the susceptibility in pure nematic 8CB 
and describes the critical thermal fluctuations on approaching 
the N-SmA transition~\cite{ocko}.  
The second term, which is simply proportional to this susceptibility squared, is designed to 
account for  static fluctuations induced by the quenched disorder.  
This term is motivated by studies on random field Ising magnets in 
which
such an expression has been shown to describe accurately the short-range correlations induced by the static 
fluctuations~\cite{RFIMexpt} and has been justified 
theoretically~\cite{grinstein,aharony}.  In the limit of 
long range order, the second term evolves into a Bragg-peak, 
$\sigma_{2}\delta(q_{\|}-q_{0})\delta(q_{\bot})$~\cite{quasi}.  The solid lines in Fig.~1 
are the results of fits to the powder average of Eq.~(1) convolved with the 
experimental resolution function.  The measured 
scattering profile also includes a temperature-independent, sloping 
background from the aerosil gel, shown in the figures as dashed 
lines.  As Fig.~1 illustrates, Eq.~(1)
provides an excellent description of the lineshape over a wide range 
of temperatures and aerosil densities.  In these fits, we have taken
$\xi_{\bot}$ and $c$ as functions of 
$\xi_{\|}$, with $\xi_{\bot}(\xi_{\|})$ and $c(\xi_{\|})$ 
set by their relations in pure 8CB, which are known to high 
precision~\cite{ocko}.  Thus, each fit has only four free parameters -- 
$q_{0}$, $\xi_{\|}$, $\sigma_{1}$, and $\sigma_{2}$ -- with two 
parameters, $\xi_{\|}$ and $\sigma_{2}/\sigma_{1}$, controlling the 
shape of the peaks.

At temperatures near and above 
the N-SmA transition temperature for pure 8CB, $T_{NA}^{0} = 306.98 
K$~\cite{germano}, the measured lineshapes 
in the aerosil samples are described very well by the thermal fluctuation term alone 
(i.e., $\sigma_{2} = 0$).  Thus, for $T > T_{NA}^{0}$, the smectic 
correlations in aerosil samples are very similar to the critical fluctuations of the pure liquid 
crystal.  Figure 2(a) displays $\xi_{\|}$ for a series of aerosil densities in this 
high temperature region.  The solid line in the figure, $\xi_{\|}$ for 
pure 8CB (aligned in the nematic by a magnetic field)~\cite{ocko}, 
demonstrates the strong similarity between thermal fluctuations in 
the pure liquid crystal and in the liquid crystal with 
dispersed aerosil.  At these high temperatures, $\xi_{\|}$ is smaller than the average void 
size, $l_{o}$, in the gels, and the influence of the gel is 
minimal.  (The values of $l_{o}$ range from 200 {\AA} for $\rho_{s}=0.341$ 
to 2650 {\AA} for 
$\rho_{s}=0.025$~\cite{germano}.)  Rather than follow the diverging behavior 
associated with a thermodynamic transition, 
however, $\xi_{\|}$ for 8CB with dispersed aerosil deviates from that of pure 
8CB as the temperature decreases, and the smectic correlations remain 
short-ranged.  Figure 2(b) shows the behavior of $\xi_{\|}$ 
for four values of $\rho_{s}$  to low temperatures.  
The susceptibility, $\sigma_{1}$, for 8CB with dispersed aerosil similarly tracks 
that of the pure system at high temperatures but fails to diverge.

We stress the importance of including the quartic term ($c>0$) in 
Eq.~(1) to account correctly 
for the temperature dependence of the lineshapes.  The values 
of $c$ vary from approximately 0.25 for 
$\xi_{\|} = 50$\AA, its value 4 K above $T_{NA}^{0}$, to less than 0.02 
for $\xi_{\|} > 10^{4}$\AA~\cite{ocko}.  At high 
temperatures, when $c \approx 0.25$, the thermal fluctuation term is effectively 
the square of a Lorentzian along $q_{\bot}$, leading 
to a powder-averaged lineshape that strongly resembles a Lorentzian.  As 
$\xi_{\bot}$ increases on 
cooling and $c$ decreases, the powder average of the first term in 
Eq.~(1) deviates strongly from 
a Lorentzian, and the tails of the lineshape become more pronounced.  We note that the x-ray 
lineshapes of 8CB in rigid aerogels resemble this 
behavior~\cite{rappaport}.  The analysis by Rappaport {\it et al.}
of the aerogel results, however, differs significantly from that presented 
here.  In particular, no measurable signal from thermal fluctuations was 
reported for 8CB in aerogel~\cite{bellini-science,rappaport}.

At temperatures below $T_{NA}^{0}$, deviations from the form describing the 
critical behavior of pure 8CB become apparent in the lineshape of 8CB with dispersed aerosil, 
particularly for large $\rho_s$.  These 
deviations indicate an increasing importance of interactions with the gel, rather than 
intrinsic thermal fluctuations, in determining the smectic correlations. 
The second term in Eq.~(1) describing random-field induced, 
short-range order accounts for these changes in the correlations 
very well.  Figure 3 displays
$\sigma_{2}$ as a function of temperature for two representative values of $\rho_s$.  
The rapid growth of $\sigma_{2}$ on cooling reflects the 
dominance of these random-field static fluctuations at low 
temperatures.  In this low 
temperature region, $\xi_{\|}$ reaches an essentially 
temperature-independent value, as Fig.~2(b)
illustrates.  Figure 4 displays the low temperature correlation 
lengths, $\xi_{\|}^{LT}$
as a function of $\rho_s$.  The solid 
line in Fig.~4 is the result of a fit to the nematic correlation lengths of the 
homolog 6CB 
with dispersed aerosils measured by Bellini 
\et ~\cite{bellini} with static light scattering.  These light 
scattering studies demonstrate that the 
nematic order in a liquid crystal with dispersed aerosil breaks up into very 
large, but finite, domains.  Since the smectic layering forms within these nematic
domains, the nematic domain sizes naturally set upper limits 
for the range of the smectic correlations.  
However, $\xi_{\|}^{LT}$ remains 
well below these limits for all $\rho_s$, demonstrating the  
sensitivity of the smectic phase to quenched disorder.  The dashed line 
in Fig.~4 is the mean aerosil 
void size, $l_{0}$, as a function of $\rho_s$~\cite{germano}.  
For the range of densities studied, the ratio of $\xi_{\|}^{LT}$ to 
$l_{0}$ is nearly independent of density and is roughly equal to four, 
consistent with an intermediate strength disordering field.  
This result contrasts with the smectic ordering in 
aerogel~\cite{bellini-science} for which the measured correlation lengths  
barely exceed the void sizes, illustrating again the weaker nature of 
disorder induced by aerosil. 
%While $\xi_{\|}^{LT}$ 
%exceeds this void size for the full range of densities, the 
%low-temperature correlated 
%volume, $(\xi_{\|}\xi_{\bot}^{2})^{LT}$, remains slightly smaller 
%than $l_{0}^{3}$ as $\rho_s$ is varied.  Thus, the  
%aerosil gels, despite their ability to anneal elastically, remain effective in disrupting the smectic 
%transition, demonstrating again the extreme fragility of the smectic phase to disorder. 

At low temperatures (where $c$ approaches a constant value) $\sigma_{2}$ 
is proportional to the integrated 
intensity of the static fluctuation term in $S(q)$.  The growth of 
this intensity strongly resembles 
that of an order parameter squared, and the solid lines 
in Fig.~3 are the results of fits to the form $\sigma_{2}\propto(T_{2}-T)^{x}$, 
where both $x$ and $T_{2}$ depend on $\rho_{s}$.  The inset of 
Fig.~3 displays the change in $x$ with $\rho_{s}$.  We emphasize that 
$x$ is an effective exponent since the aerosil destroys the true N-SmA 
phase transition.  
%The strength of this coupling 
%depends on the magnitude of the nematic susceptibility at $T_{NA}$, 
%which for pure liquid crystals depends on the width of the nematic 
%range and can be roughly parameterized by the MacMillan ratio, 
%$R_{M} \equiv T_{NA}/T_{NI}$.  In the limit of zero 
%coupling ($R_{M}\rightarrow 0$) the N-SmA transition is believed to 
%exhibit 3D-XY universality, while it approaches tricritical behavior and 
%can even become first order as 
%$R_{M}\rightarrow 1$~\cite{nounesis}.  Values of $2\beta$, 
%determined through the scaling relation $2\beta = 2-\alpha-\gamma$~\cite{beta}, for a 
%series of liquid crystals~\cite{nounesis} with varying $R_{M}$ are included in the 
%inset to Fig.~3.  
However, associating the integrated 
intensity from the static, short-range correlations with an order 
parameter squared (so that $x=2\beta$), we observe 
that the effective exponent spans the range from near the tricritical value (like with 
pure 8CB) at small $\rho_{s}$ to 3D XY-like at large $\rho_{s}$.  For pure liquid crystals,
the measured critical behavior at the N-SmA transition is affected by 
couplings between 
the nematic and smectic order parameters
(de Gennes coupling)~\cite{nounesis}.  The variation of $x$ with $\rho_{s}$
suggests that the quenched disorder suppresses the nematic susceptibility 
that drives the critical behavior 
of the pure system away from 3D-XY universality.  This result agrees  
nicely with specific heat studies of 8CB with dispersed 
aerosil which show the effective heat capacity exponent, $\alpha$, 
decreasing toward the 3D-XY value with increasing 
$\rho_{s}$~\cite{germano,fss}.  

In conclusion, these x-ray scattering studies have shown that 8CB 
with dispersed aerosil provides a model random field system for a transition that breaks a 
continuous symmetry.  In particular, the success of Eq.~(1) in describing the 
smectic correlations and the corresponding behaviors 
of $\xi_{\|}$ and $\sigma_{2}$ form a compelling picture in which the 
aerosil gel creates a random local-orienting field that destroys the N-SmA 
transition, consistent with theoretical 
predictions~\cite{leo}.  In the random field Ising magnet, slow dynamics 
and non-equilibrium effects strongly influence the 
behavior~\cite{fisher,feng}.  Efforts 
to observe hysteretic or time dependent behavior indicative of 
metastability in these scattering measurements or in the calorimetry 
studies~\cite{germano} on 8CB with 
dispersed aerosil revealed no such effects.  
These results suggest that the continuous symmetry allows the system to find equilibrium 
efficiently, even in the presence of random fields, in marked contrast 
to the behavior in random field Ising systems~\cite{birgeneau}.
However, a better understanding of the 
dynamics would be beneficial for placing the observed ordering in the context of 
equilibrium statistical mechanics.

We thank S. Lamarra for technical support and T. Bellini, P. Clegg and 
L. Radzihovsky for helpful discussions.  
This work was supported by the NSF under Contract No.~DMR-0071256 (MIT) 
and NSF-CAREER Award No.~DMR-0092786 (WPI) and by the Natural Science and 
Engineering Research Council of Canada (Toronto).  J.-L. G. acknowledges support 
from CNRS and NATO (grant 31C96FR).

%%%%%%%%%%%%%%%%%%%%%%%%%%%%%%%%%%%%%%%%%%%%%%%%%%%%%%%%%%%%%%%%%%%%%%%%

\begin{figure}
\caption{X-ray scattering intensity for 8CB with dispersed aerosil for 
$\rho_{s}=0.025$ g/cm$^{3}$ at (a) 288.5 K and (b) 307.6 K 
and $\rho_{s}=0.282$ g/cm$^{3}$ at (c) 300.6 K and (d) 306.6 K.  The solid lines 
are the results of fits to the powder average of Eq.~(1) convolved with 
the experimental
resolution and the dashed lines are the temperature independent 
background due to scattering from the aerosil gel.  The higher 
temperature lineshapes, (b) and (d), are due to thermal critical fluctuations 
as in pure 8CB while the low temperature lineshapes, (a) and (c), have 
forms expected for short-range correlations dominated by static, random-field induced fluctuations.}
\label{fig1}
\end{figure}

\begin{figure}
\caption{The parallel correlation length, $\xi_{\|}$, as a function 
of temperature (a) at high temperature for a series of aerosil densities:  
$\rho_{s}=0.025$ ($\bigcirc$), $0.041$ ($\bullet$), $0.051$ (filled 
diamond),
$0.078$ (open box), $0.105$ ($\nabla$), $0.161$ ($\triangle$),
$0.220$ (filled box), $0.282$ (filled triangle), and $0.341$ 
(filled upside-down triangle); and (b) on an expanded scale to low 
temperature for four values of $\rho_{s}$.  The solid line is $\xi_{\|}$ for pure 
8CB [14].}
\label{fig2}
\end{figure}

\begin{figure}
\caption{The integrated intensity of the static fluctuation term, 
$\sigma_{2}$, as a function of temperature for 
$\rho_{s}=0.025$ ($\bullet$) and $\rho_{s}=0.220$ ($\bigcirc$).  The 
absolute magnitudes of $\sigma_{2}$ for each $\rho_{s}$ are arbitrary
and should not be compared.  The solid lines 
are fits to $\sigma_{2}\propto(T_{2}-T)^{x}$.  The inset shows $x$ 
versus $\rho_{s}$.  The
effective exponent $x$ spans the 
range for an order parameter exponent from near the tricritical value, $0.5$, at small $\rho_{s}$ to 
the 3D XY value, $0.69$, at large $\rho_{s}$.}
\label{fig3}
\end{figure}

\begin{figure}
\caption{Low temperature correlation lengths, $\xi_{\|}^{LT}$ versus $\rho_{s}$.  
The solid line is the nematic 
correlation length (for 6CB) [19], and the 
dashed line is the mean aerosil void size, $l_{0}$.}
\label{fig4}
\end{figure}


\begin{references}
\bibitem{quasi}
Due to the Landau-Peierls instability the correlations in the smectic 
phase exhibit algebraic order rather than true long-range 
order.  This distinction should not influence the form of the correlations 
in the presence of quenched disorder.
\end{references}
\end{document}